\documentclass{emulateapj}
\usepackage{amsmath}

\def\stacksymbols #1#2#3#4{\def\theguybelow{#2}
        \def\verticalposition{\lower#3pt}
        \def\spacingwithinsymbol{\baselineskip0pt\lineskip#4pt}
        \mathrel{\mathpalette\intermediary#1}}
\def\intermediary #1#2{\verticalposition\vbox{\spacingwithinsymbol
        \everycr={}\tabskip0pt
        \halign{$\mathsurround0pt#1\hfil##\hfil$\crcr#2\crcr
                \theguybelow\crcr}}}
\def\lta{\stacksymbols{<}{\sim}{2.5}{.2}}

\shorttitle{X-RAY SUPERCAVITIES AND THEIR IMPACT ON GALAXY CLUSTERS}
\shortauthors{GUO \& MATHEWS}

\begin{document}
\bibliographystyle{apj} 

\title{Simulating X-ray Supercavities and Their Impact on Galaxy Clusters } 
\author{Fulai Guo\altaffilmark{1} and William G. Mathews \altaffilmark{1}}

\altaffiltext{1}{UCO/Lick Observatory, Department of Astronomy and Astrophysics, University of California, Santa Cruz, CA 95064, USA; fulai@ucolick.org}

\begin{abstract}
Recent X-ray observations of hot gas in the galaxy cluster MS 0735.6+7421
reveal huge radio-bright, quasi-bipolar X-ray cavities having a total
energy $\sim 10^{62}$ ergs, the most energetic AGN outburst currently
known. We investigate the evolution of this outburst with
two-dimensional axisymmetric gasdynamical calculations in which the
cavities are inflated by relativistic cosmic rays.  Many key
observational features of the cavities and associated shocks are
successfully reproduced.  The radial elongation of the cavities
indicates that cosmic rays were injected into the cluster gas by a
(jet) source moving out from the central AGN. 
AGN jets of this magnitude must be almost perfectly identically bipolar. 
The relativistic momentum of a single jet
would cause a central AGN black hole of mass $10^9$ $M_{\odot}$ to
recoil at $\sim 6000$ km s$^{-1}$, exceeding kick velocities during 
black hole mergers, and be ejected from the cluster-center galaxy.
Observed deviations from bipolar symmetry in the radio cavities can be caused 
by subsonic flows in the ambient cluster gas, but reflection shocks between 
symmetric cavities are likely to be visible in deep X-ray images.
When the cavity inflation is complete,
$4PV$ underestimates the total energy received by the cluster gas.
Deviations of the cluster gas from hydrostatic equilibrium are most pronounced
during the early cavity evolution when the integrated cluster mass
found from the observed gas pressure gradient can have systematic
errors near the cavities of $\sim$10-30\%.  
The creation of the cavity with cosmic rays 
generates a long-lasting global cluster expansion that
reduces the total gas thermal energy
below that received from the cavity shock -- even this most energetic
AGN event has a net cooling effect on cluster gas.  
One Gyr after this single outburst, 
a gas mass of $\sim 6 \times 10^{11}$ $M_{\odot}$ 
is transported out beyond a cluster radius of 500 kpc.
Such post-cavity outflows can naturally produce the discrepancy observed between the
cluster gas mass fraction and the universal baryon fraction inferred
from WMAP observations.

\end{abstract}

\keywords{
cosmic rays-- cooling flows -- galaxies: clusters: general -- galaxies: active -- X-rays: galaxies: clusters}

\section{Introduction}
\label{section:intro}

The hot gas in galaxy clusters emits prolifically in X-rays, and has been extensively studied by X-ray telescopes {\it Chandra} and {\it XMM-Newton}, revealing complex structures such as X-ray cavities, thermal filaments, and weak shocks in the intracluster medium (ICM) as in the Perseus cluster (\citealt{fabian06}). These interesting structures are presumed to be related to the active galactic nuclei (AGNs) located at the cluster centers. 
Although the ICM in many clusters losses energy by radiation in a timescale much shorter than the cluster age, high-resolution X-ray spectroscopy indicates that the gas does not cool to low temperatures (e.g., \citealt{2001A&A...365L.104P, 2003ApJ...590..207P, 2001A&A...365L..87T}; for a review see
\citealt{2006PhR...427....1P}). The deficit of cool-phase gas 
and young stars in cluster cores, often known as the `cooling flow' problem, suggests that the ICM is prevented from the cooling catastrophe by one or more heating sources. Currently, heating of the ICM by central AGNs, as indicated by X-ray cavities and weak shocks, is regarded as the most successful energy source to balance radiative cooling (see \citealt{2007ARA&A..45..117M} for a recent review). Recent observations show that the AGN energies associated with X-ray cavities ($\sim 4PV$, where $P$ is the average pressure of the gas surrounding the cavity, and $V$ is the cavity's volume) are sufficient or nearly sufficient to stop the gas from cooling in most clusters containing detectable cavities \citep{rafferty06}. Furthermore, the mean power of these AGN outbursts increases in proportion to the cluster cooling luminosity (\citealt{birzan04}; \citealt{rafferty06}). This correlation suggests
that AGN outbursts are triggered by the gas cooling and operates as a
self-regulating feedback mechanism, which is shown to be essential in
suppressing global thermal instability and thus in maintaining the ICM
in the cool core state \citep{guo08b}.

Despite these supportive clues, the detailed mechanisms that transfer
the AGN jet energy to the thermal energy of the ICM are still far from
clear. Numerical simulations have been extensively used to understand
this process. X-ray cavities are usually thought to form at the tips
of AGN-induced jets. During cavity formation, weak shock waves
emerge as the gas is displaced, and heat the ICM 
\citep{brueggen07}. The buoyantly-rising cavities may induce sound waves, which may be viscously dissipated in the ICM and thus heat the gas
(\citealt{ruszkowski04}; but see \citealt{mathews06}).
Radio synchrotron emission has been observed
from many X-ray cavities \citep{birzan04}, suggesting that the
cavities are formed by AGN-induced cosmic rays (CRs), which may
leak into the ICM and heat the gas \citep{2008MNRAS.384..251G}.

In most previous simulations, for simplicity, it is assumed that X-ray cavities
are filled with ultrahot thermal gas, but CR-filled cavities have 
received much less attention. The only explicit studies of the
formation and evolution of X-ray cavities filled with CRs are
\citet{mathews08a}, \citet{mathews08}, and \citet{mathews09}, who show several interesting
effects of CR-formed X-ray cavities: ICM heating due to weak shock
generation, ICM cooling associated with cluster
expansion and outward mass transfer, and 
the dynamical relationship between post-cavity X-ray filaments 
and extended radio lobes. However, these studies focus on low-energy cavities ($\sim
10^{58}-10^{59}$ erg) in the Virgo cluster. In this paper, we apply
the same model to study 
the remarkably large X-ray supercavities observed in the cluster MS
0735.6$+$7421 (hereafter MS0735), created by the most energetic AGN
outburst known \citep{mcnamara05, gitti07, mcnamara2009}. In this cluster, two
approximately symmetric cavities are observed in X-ray images, each
associated with relativistic gas emitting radio synchrotron
emission. The energy released by this outburst is estimated to be
$6.4\times 10^{61}$ erg, assuming an enthalpy of $4pV$ per cavity
\citep{rafferty06}. Weak shocks have also been detected in MS0735; a
simple spherically-symmetric shock model suggests that the age and
driving energy of the shock are $t_{\rm s}=104$ Myr and $E _{\rm
  s}=5.7\times 10^{61}$ erg, respectively \citep{mcnamara05}.
 MS0735 has been studied numerically by \citet{soker09}, who assume that 
 the cavities are formed by slow, massive, wide jets with a total injected energy of 
 $7.2\times 10^{61}$ erg and a very large injected mass $\sim 10^{10}M_{\sun}$.
 
One of the main objectives of the current 
paper is to investigate if AGN
bubbles created by cosmic rays can reproduce the observed morphology
of giant X-ray supercavities in X-ray maps of MS0735. We show that the
cavity morphology contains important information about the process of
CR injection: CRs are likely to have been injected into the ICM at a series of
locations (or continuously) along the direction of a jet as it moves
outward, instead of at a fixed location
(\S~\ref{section:morphology}). We then follow the time evolution of
the ICM energies during and after cavity creation and study the
energetics of X-ray supercavities and their long-term influence on the cluster gas atmosphere. 
The investigation of this extremely powerful AGN outburst improves our understanding of how
AGN outbursts affect the ICM, and their relevance in solving the cooling flow
problem.

As the largest virialized structures in the universe, clusters of
galaxies are useful cosmological probes, providing one of the current
best constraints on the mean matter density $\Omega_{\rm m}$, dark
energy density $\Omega_{\rm DE}$, and the dark energy equation of
state parameter $w$ (see \citealt{allen08} and references
therein). These constraints are usually based on accurate measurements
of the X-ray gas mass fraction profile $f_{\rm gas}(r)$. In this paper, we study
how the energetic AGN outburst in MS0735 affects the gas mass fraction
by producing a buoyant outflow and global expansion of the cluster thermal gas. X-ray
supercavities also have a significant effect on the assumptions of
hydrostatic equilibrium and spherical symmetry, which are essential in
measuring the cluster's total mass profile from X-ray observations.
 
The rest of the paper is organized as follows. In
\S~\ref{section:equations}, we describe basic time-dependent equations
and our numerical setup. Our results are presented in
\S~\ref{section:results}. We summarize our main results in
\S~\ref{section:conclusion} with a discussion of the implications.

\section{Equations and Numerical Setup}
\label{section:equations}

To study the dynamical effects of CRs on the hot cluster gas 
following this extremely powerful AGN outburst in MS0735,
we assume that AGN outbursts inject cosmic rays into the ICM, producing the observed
radio lobes, X-ray supercavities and weak shocks \citep{mcnamara05,
  gitti07,mcnamara2009}. The combined evolution of the thermal ICM and
cosmic rays may be described by the following four equations:

\begin{eqnarray}
\frac{d \rho}{d t} + \rho \nabla \cdot {\bf v} = 0,\label{hydro1}
\end{eqnarray}
\begin{eqnarray}
\rho \frac{d {\bf v}}{d t} = -\nabla (P+P_{\rm c})-\rho \nabla \Phi ,\label{hydro2}
\end{eqnarray}
\begin{eqnarray}
\frac{\partial e}{\partial t} +\nabla \cdot(e{\bf v})=-P\nabla \cdot {\bf v}
   \rm{ ,}\label{hydro3}
   \end{eqnarray}
\begin{eqnarray}
\frac{\partial e_{\rm c}}{\partial t} +\nabla \cdot(e_{\rm c}{\bf v})=-P_{\rm c}\nabla \cdot {\bf v}+\nabla \cdot(\kappa\nabla e_{\rm c})+\dot{S_{\rm c}}
   \rm{ ,}\label{hydro4}\\ \nonumber
   \end{eqnarray}
\noindent
where $d/dt \equiv \partial/\partial t+{\bf v} \cdot \nabla $ is the
Lagrangian time derivative, $P_{\rm c}$ is the CR pressure, $e_{\rm
  c}$ is the CR energy density, $\kappa$ is the CR diffusion
coefficient, $\dot{S_{\rm c}}$ is the CR source term due to the
central AGN activity, and all other variables have their usual
meanings. Pressures and energy densities are related via
$P=(\gamma-1)e$ and $P_{\rm c}=(\gamma_{\rm c}-1)e_{\rm c}$, where we
assume $\gamma=5/3$ and $\gamma_{\rm c}=4/3$.  
We do not include radiative cooling since our intention is to 
study the evolution of X-ray supercavities in the cluster MS0735 
on timescales short compared to the multi-Gyr age of the cluster, 
i.e. we do not investigate  
the long-term balance between radiative cooling and intermittent heating by 
AGN outbursts. The same set of equations and underlying physical
assumptions have been thoroughly described in \citet{mathews08} and
\citet{mathews09} where further details can be found. 
Here we simply summarize several modifications and reiterate
some important points. 

Equations (\ref{hydro1}) $-$ (\ref{hydro4}) are solved in $(r, z)$
cylindrical coordinates using a two-dimensional Eulerian code similar
to ZEUS 2D \citep{stone92}; in particular, we have incorporated into
the code a background gravitational potential, CR diffusion, and CR
energy equation. The computational grid consists of $200$ equally
spaced zones in both coordinates out to $400$ kpc plus an additional
$200$ logarithmically-spaced zones out to $2$ Mpc. 
In view of this large computational domain, we adopt reflective
boundary conditions for thermal gas and cosmic rays at 
the outer boundary as well as at the origin.

For initial conditions, we adopt analytic fits to the deprojected electron number
density $n_{\rm e}$ and temperature $T$ profiles observed by \citet{gitti07}:
\begin{eqnarray}
n_{\rm e}(r)=\frac{0.075}{[1+(r/20)^{2}]^{1.29}}+\frac{0.01}{[1+(r/200)^{2}]^{1.15}} {\rm ~cm}^{-3}
\end{eqnarray}
and
\begin{eqnarray}
T(r)=(T_{1}^{-1.5}+T_{2}^{-1.5})^{-2/3}{\rm ~keV,}
\end{eqnarray}
where 
\begin{eqnarray}
T_{1}=3.2+5.3(r/275)^{1.7} {\rm ~,}
\end{eqnarray}
\begin{eqnarray}
T_{2}=8.5(r/275)^{-0.7} {\rm ~,}
\end{eqnarray}
and $r$ is measured in kpc. The fits are shown as solid lines in Figure \ref{plot4} (a) and (b). At the
beginning of our simulation, the CR energy density is
assumed to be zero throughout the cluster. The gravitational potential
$\Phi$ is set by assuming hydrostatic equilibrium at $t=0$.

X-ray cavities are usually thought to be inflated by bipolar jets
emanating from an AGN in the central galaxy. 
The approximately symmetric double-cavity morphology in MS0735 
suggests that the outburst in this cluster was nearly 
bipolar-symmetric, as we assume here.
The jets deposit relativistic cosmic
rays into small regions at their terminal points, which expand and
form underdense bubbles producing the observed X-ray cavities.
The injection of cosmic rays into the ICM is described 
in equation (\ref{hydro4}) by the
source term $\dot{S_{\rm c}}$. Similar to
\citet{mathews08} and \citet{mathews09}, we assume that the CRs are
deposited into a Gaussian-shaped sphere of characteristic radius
$r_{\rm s}=2$ kpc located at ${\bf r}_{\rm cav}=(r,z)=(0, z_{\rm
  cav})$:
\begin{eqnarray}
\dot{S_{\rm c}}= 
\begin{cases}
\frac{E_{\rm agn}}{t_{\rm agn}}\frac{e^{-[({\bf r}-{\bf r}_{\rm cav})/r_{\rm s}]^{2}}}{\pi^{3/2}r_{\rm s}^{3}}      & \quad \text{when $t \leq t_{\rm agn}$,}\\ 
0& \quad \text{when $t>t_{\rm agn}$,}
\end{cases}
\end{eqnarray}
\noindent
where $t_{\rm agn}$ is the duration of the CR injection (AGN active
phase), and $E_{\rm agn}$ is the total injected CR energy in one
bubble ($2E_{\rm agn}$ for the whole cluster). The integral of
$\dot{S_{\rm c}}$ over space gives the the CR injection luminosity
$E_{\rm agn}/t_{\rm agn}$ associated with the creation of each
bubble. To mimic the location and size of the observed X-ray cavity,
we consider three cases for the location of the CR injection ${\bf
  r}_{\rm cav}$: (a) $z_{\rm cav}$ is fixed at $60$
kpc (adopted in run MS-1); (b) $z_{\rm cav}$ moves from $40$ to $160$ kpc 
at a constant speed during the CR injection phase (adopted in runs MS-2 and MS-2A); 
and (c) $z_{\rm cav}$ moves from $40$ to $160$
kpc with a constant deceleration during the CR injection phase (the
speed drops to zero at $z_{\rm cav}=160$ kpc; adopted in run MS-3). See Table \ref{table1}
for specific model parameters in each run.
The bulk kinetic energy of cosmic rays is negligible compared to the 
injected cosmic ray energy $E_{\rm agn}$ due to their negligible total rest mass.
Here we assume that the energy that AGN jets deposit into the ICM is mainly in
the form of cosmic rays, which are transported along the jets and created in strong shocks as the jets encounter the ICM. However, the energy content of real AGN jets is not clear and may also 
include magnetic energy, the internal and kinetic energies of thermal (and superthermal) gas. 
An MHD jet simulation including cosmic ray acceleration (and re-acceleration) mechanisms 
is required to fully understand this process.

In addition to their advection with the thermal gas, cosmic rays
diffuse through the gas as 
described in equation (\ref{hydro4}). The CR diffusion coefficient
$\kappa$ is poorly known but may vary inversely with the gas
density since the magnetic field is probably larger in denser gas
\citep{dolag01}. As in \citet{mathews08} and \citet{mathews09}, 
we adopt the following functional dependence of the diffusion
coefficient on the gas density:
\begin{eqnarray}
\kappa=
\begin{cases}
10^{30}(n_{{\rm e}0}/n_{\rm e}) \text{~cm}^{2} \text{~s}^{-1}    & \quad \text{when } n_{\rm e}> n_{{\rm e}0} \text{,}\\
10^{30} \text{~cm}^{2} \text{~s}^{-1}    & \quad \text{when } n_{\rm e} \leq n_{{\rm e}0} \text{,}
\end{cases}
\end{eqnarray}
\noindent
where we take $n_{{\rm e}0}=10^{-5}$ cm$^{-3}$ in the rest of the
paper. Models with different values of $n_{{\rm e}0}$ have been
explored by \citet{mathews08} and the results are usually not very
sensitive to it. During their diffusion, cosmic rays interact with
magnetic irregularities and Alfv\'{e}n waves, exerting CR pressure
gradients on the thermal gas (equation \ref{hydro2}). 
The shock generation and early evolution of cavities are mainly caused by 
the expansion of the surrounding gas driven by the CR pressure unless the coefficient of CR diffusion is so large (e.g., $n_{{\rm e}0}\gtrsim 10^{-2}$ cm$^{-3}$) that it significantly affects 
the morphology of cavities. We neglect other
more complicated interactions of cosmic rays with thermal gas, e.g.,
Coulomb interactions, hadronic collisions, and
hydromagnetic-wave-mediated CR heating, which depend on the cosmic ray
energy spectrum and provide additional heating effects for the ICM
(e.g., \citealt{2008MNRAS.384..251G}). We defer a thorough study of
these effects to future work.

\begin{table}
 \centering
 \begin{minipage}{70mm}
  \renewcommand{\thefootnote}{\thempfootnote} 
  \caption{List of Simulations.}
    \vspace{0.1in}
  \begin{tabular}{@{}lccccc}
  \hline & {$E_{\rm agn}$\footnote{The AGN energy released in the form
      of cosmic rays in one hemisphere during the outburst.
      } }&
         {$t_{\rm agn}$\footnote{The duration of the active AGN
             phase.}} & {${\bf r}_{\rm cav}$\footnote{${\bf r}_{\rm
               cav}=(0, z_{\rm cav}$) is the position where the cosmic
             rays are injected; see text in \S~\ref{section:equations}
             for three different ways of cosmic ray injection.}} &
         {$t_{\rm shock}$\footnote{The time that the shock reaches
             $r=240$ kpc along the semi-minor axis (i.e., the age of the energetic
outburst).}} & {$M_{\rm
             shock}$\footnote{The Mach number of the shock along the
             semi-minor axis when it reaches $r=240$ kpc.}}\\ Run&
         $(10^{61}$ erg)& (Myr)&&($10^{8}$ yr)&\\ \hline MS-1 &4.5&10
         &a& 1.10 &1.41 \\ MS-2 & 3.8& 10 &b& 1.16&1.41 \\ MS-3 & 3.5&
         10 &c&1.23 &1.41 \\ MS-2A &3.8 & 50 &b& 1.38 &1.34 \\ \hline
\label{table1}
\end{tabular}
\end{minipage}
\end{table}

\section{Results}
\label{section:results}

\begin{figure*}
\plotone{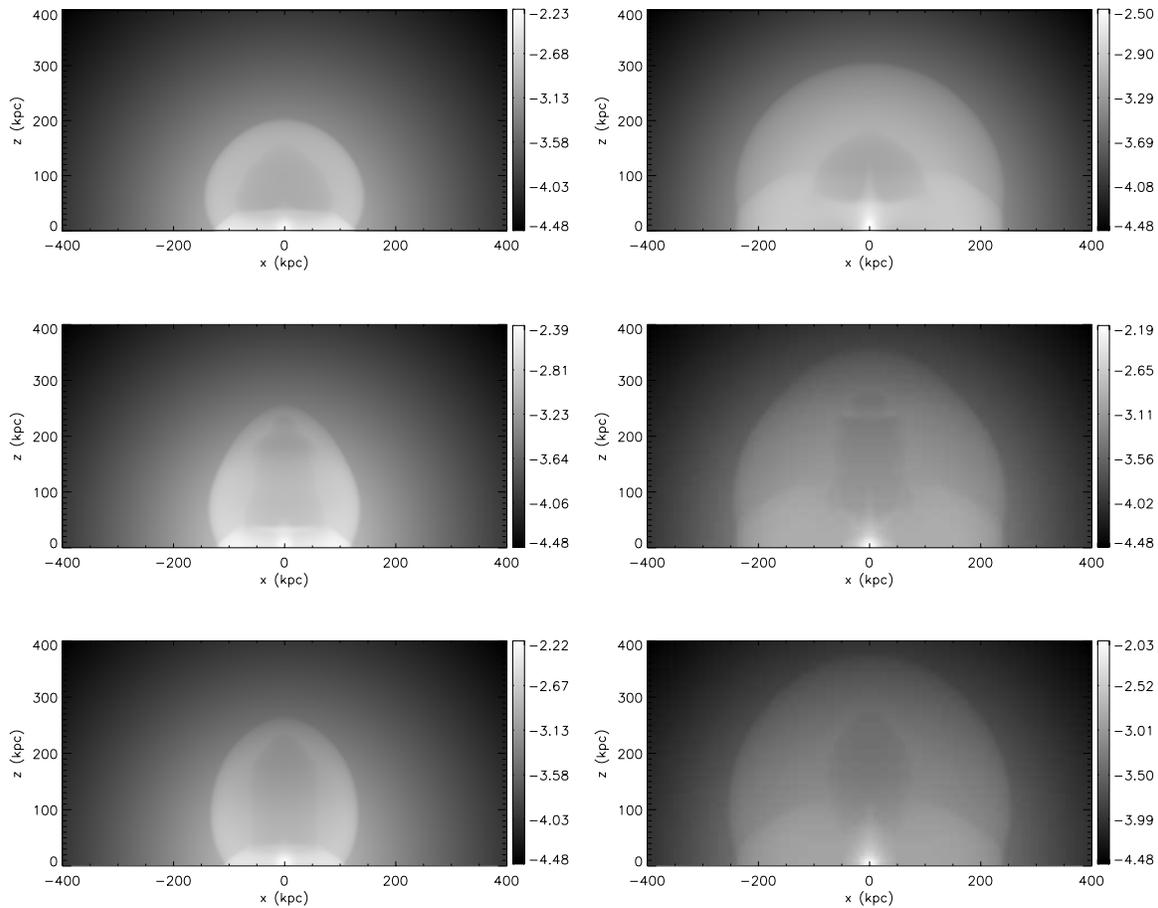}
\caption{Synthetic X-ray surface brightness maps (projections of
  $n_{\rm e}n_{\rm i}\Lambda(T,Z)$ viewed perpendicular to the bubble 
  symmetry axis)
  for runs MS-1 ({\it top}), MS-2 ({\it middle}), and MS-3 ({\it
    bottom}) in cgs units with logarithmic scale. 
    Left column: $t=50$ Myr; right column: $t=t_{\rm
    shock}$ (see Table \ref{table1}). X-ray supercavities are clearly
  visible. The prolate shapes of the cavities in runs MS-2 and MS-3
  fit observations \citep{mcnamara2009} better than the oblate cavity
  seen in run MS-1.  }
 \label{plot1}
 \end{figure*}

\subsection{Morphology and Structure of X-ray Supercavities}
\label{section:morphology}

{\it Chandra} observations show that the cluster MS0735 hosts two
giant cavities. Each cavity is slightly elongated in the radial
direction (`prolate') and is roughly $200$ kpc in (radial) diameter
\citep{mcnamara05, mcnamara2009}. Weak shocks have also been detected;
the shock front is roughly elliptical with a semimajor axis of $\sim
360$ kpc and a semiminor axis of $\sim 240$ kpc \citep{gitti07}. Along
the semiminor axis (perpendicular to the cavity axis), 
the shock Mach number is estimated to be $\sim
1.41$ \citep{mcnamara05}.

One of the primary goals of this study is to investigate whether AGN
bubbles created by cosmic rays can reproduce these observational
features. Adopting the radiative cooling function $n_{\rm e}n_{\rm
  i}\Lambda(T,Z)$ from \citet{1993ApJS...88..253S}, we project the
emissivity and produce synthetic X-ray maps of our runs at different
times. Here $n_{\rm i}$ is the total number density of ions and we
take an average metallicity of $Z=0.4$ (see \citealt{gitti07}). 
$n_{\rm i}$ is related to the proton number density $n_{\rm p}$
via $n_{\rm i}=1.1n_{\rm p}$, and thus the molecular weight is $\mu=0.61$. 
Figure (\ref{plot1}) shows synthetic X-ray surface brightness maps in
logarithmic scale for our three typical models (MS-1, MS-2, MS-3) as
listed in Table \ref{table1} at two different times $t=50$ Myr ({\it
  left panels}) and $t=t_{\rm shock}$ ({\it right panels}), where
$t_{\rm shock}$ is the time when the shock propagates to $r=240$ kpc
along the semiminor axis, determining the age and energy of the 
outburst. X-ray deficient cavities and shock fronts are clearly seen
in all these runs, which differ mainly in the location of cosmic ray
injection. The CR injection time in all these runs is assumed to be
$10$ Myr, and the injected CR energy $E_{\rm agn}$ is determined so
that the shock along the semiminor axis has a Mach number of $1.41$
when propagating to $r=240$ kpc. 

In run MS$-1$, cosmic rays are
injected at a fixed position ($z_{\rm cav}=60$ kpc in the $z$
axis). The {\it top} panels of Figure (\ref{plot1}) show that the 
X-ray cavity has a roughly
oblate shape. However, the observed cavity is
elongated along the radial direction, which suggests that CRs may be
injected into the ICM at a range of $z_{\rm cav}$ as the jet tip moves
outward. Thus, in runs MS-2 and MS-3, we assume that $z_{\rm cav}$
moves from $40$ to $160$ kpc within $t \leq t_{\rm agn}$ (see
\S~\ref{section:equations} for details). The {\it middle} and {\it
  bottom} panels clearly show radially prolate cavities, consistent with
observations. Our methods of CR injection are probably idealized, but
the resulting cavity shapes strongly suggest that CRs may indeed
be injected into the ICM at a series of locations (or continuously) as
AGN jets move outward, instead of at a fixed location as assumed by
most authors (e.g., \citealt{ruszkowski07}; \citealt{brueggen09a}).

Images of the cluster MS0735 combining both X-ray and radio
wavelengths indicate that the X-ray cavities are filled with radio
emission, as seen in Fig. 1 of \citet{mcnamara2009}. 
While the overall radio image in MS0735 has an approximate north-south 
bipolar symmetry relative to the cluster-centered galaxy, 
several deviations from perfect symmetry are apparent. 
Weak radio emission observed within about 50-70 kpc from the center 
has a noticeably different alignment. It is possible that the 
inner radio structure has been moved from an original north-south alignment by 
flows in the hot gas or that MS0735 has ejected radio plasma 
in several different directions. Most of the bright extended radio emission comes from radially elongated elliptical-shaped regions that are well separated from the cluster
center. Considering a typical synchrotron-loss
timescale of less than about $100$ Myr (depending on both the magnetic field
strength and electron Lorentz factor) for relativistic electrons in
clusters, the decline in radio flux close to the center of MS0735 
could arise if the radio electrons deposited there have radiated most of 
their initial energy. This is consistent with our result 
that CRs are injected into the ICM at a series of
locations (or continuously) as AGN jets move outward. Since electrons
with lower energies have longer synchrotron-loss times and emit in
longer wavelengths, future radio observations at 
different wavelengths could provide further insights regarding 
the formation and evolution of the MS0735 X-ray cavities.
 
 \begin{figure*}
\plotone{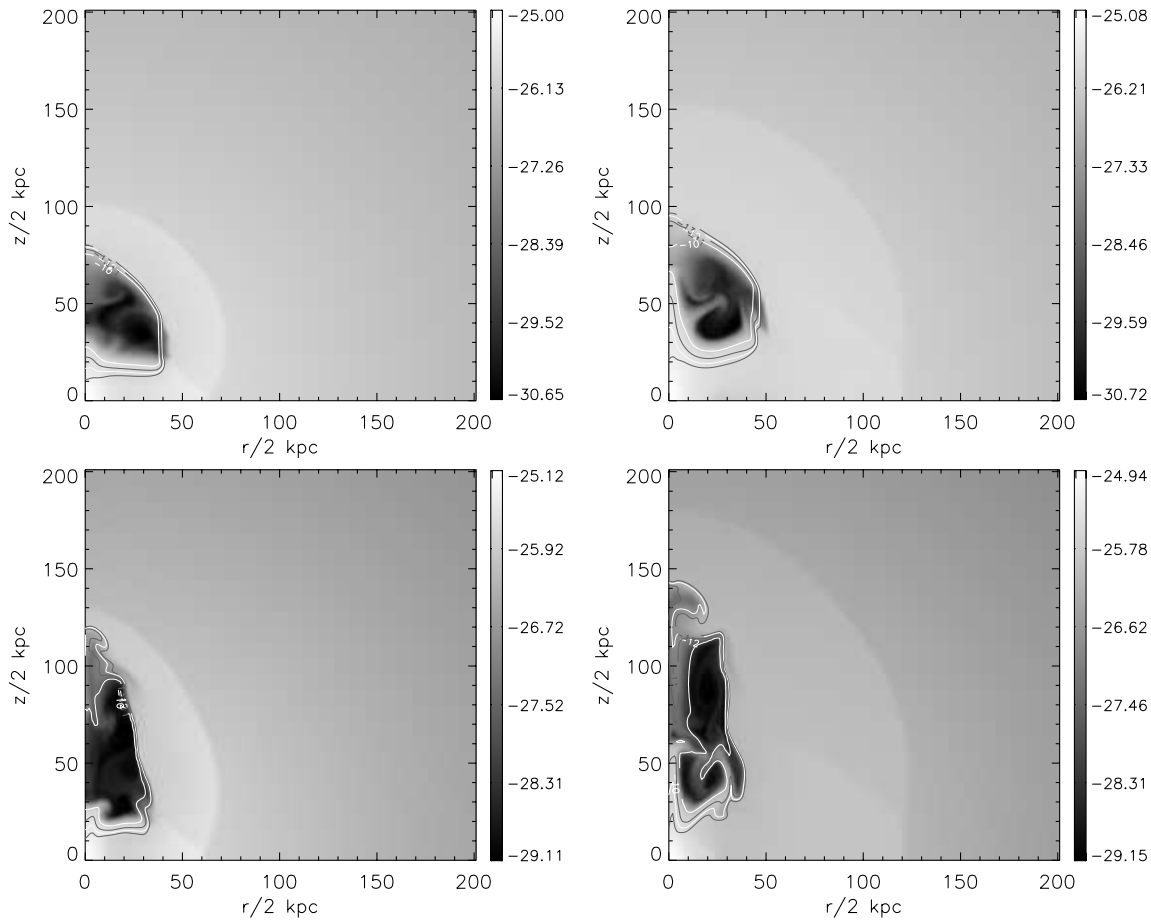}
\caption{Slices of log $(\rho/{\rm g~ cm}^{-3})$ in runs MS-1 ({\it
    top}) and MS-2 ({\it bottom}). Left column: $t=50$ Myr; right column:  $t=t_{\rm
    shock}$ (see Table \ref{table1}). White contours showing log
  $(e_{\rm cr}/{\rm erg~cm}^{-3})$ indicate that cosmic rays
  concentrate within the X-ray supercavities.
}
 \label{plot2}
 \end{figure*}

Slices of logarithmic gas density through the cluster center are shown
in Figure \ref{plot2} for runs MS-1 ({\it top panels}) and MS-2 ({\it
  bottom panels}) at $t=50$ Myr ({\it left}) and $t=t_{\rm shock}$
({\it right}). Contours show the logarithmic CR energy density. The
bubbles are clearly seen as low-density regions along $z$ axis, where
cosmic rays also concentrate. In run MS-2, where CRs are injected from
$z_{\rm cav}=40$ to $160$ kpc along $z$ axis, the bubble structure at
$t=t_{\rm shock}$ is much more complex than that in run MS-1, although
the cavities in the projected X-ray surface brightness maps are simple
and smooth in both cases. Since the prolate cavity in run MS-2 fits
observation better, this raises the possibility that the cavities in
X-ray maps may have a complex structure, with large variations of gas
density and CR energy density at different locations inside cavities. 
Also of interest is the buoyant flow of the CR cavities. 
At the time of observation $t_{\rm shock}$ the dark cavities in Figure 2 have moved well 
away from the CR source regions, $z/2 = 30$ kpc for MS-1 and 
$20 < z/2 < 80$ kpc for MS-2.

In addition to the shock confirmed by observation, another shock 
can be seen in each panel of Figure \ref{plot1} and on closer inspection 
also in Figure \ref{plot2}. This second shock,
obviously due to the reflection of the first shock at the 
symmetry boundary $z=0$, represents the shock produced by the other 
identical cavity in the cluster at $z < 0$.
It goes through the already shocked gas and has a much
smaller Mach number (with a maximum value of $\sim 1.1$ far from
the X-ray cavity). Observational efforts to detect this second shock
(even non-detection) could provide interesting information on
the geometry and simultaneity of bipolar jets and cavities in clusters.

\subsection{Outburst Momentum and Black Hole Recoil}

As the most powerful known AGN outburst, MS0735 highlights another 
fundamental concern regarding the joint evolution of massive elliptical 
galaxies and their central supermassive black holes -- why the masses
of the black holes stellar bulges are so strongly correlated 
(\citealt{ferrarese00};  \citealt{gebhardt00}). Since elliptical galaxies are commonly regarded as the end products of galaxy mergers, the black hole-stellar mass correlation suggests that 
the black holes merge in the same fashion as the surrounding stars.
However,  detailed studies show that asymmetric gravitational radiation 
emitted during the final stages of black hole mergers results in recoil black hole velocities
$\lta 4000$ km s$^{-1}$ \citep{campanelli07} sufficient to displace them far from the galactic cores 
\citep{madau04}. Nevertheless, dedicated searches for peculiar AGN recoil velocities 
or binary AGNs have resulted in surprisingly few candidates.

However, active, jet-emitting black holes 
in powerful radio galaxies like MS0735 could also receive 
very large recoil kicks if the jets are not exactly balanced 
by an identical counterjet.
Suppose for example that the total energy 
$M_{\rm cav}c^2 = 3.8 \times 10^{61}$ ergs contained 
in each large cavity of MS0735 is purely relativistic with 
a mass equivalent of $M_{\rm cav} = E_{\rm agn}/c^2 = 2.1 \times 10^7$ $M_{\odot}$.  
If this energy was supplied to the cavity by a relativistic jet  
as observations indicate, its momentum 
$p = M_{\rm cav}c = 1.26 \times 10^{51}$ 
gm cm s$^{-1}$ is sufficient to produce a recoil velocity 
of $6300$ km s$^{-1}$ for a black hole of mass $M_{\rm bh} = 10^9$ $M_{\odot}$. 
This velocity is large enough to completely eject the massive 
black hole from typical group and cluster-centered elliptical galaxies \citep{merritt04}.
To avoid this undesired fate, it is necessary that 
the momentum of the MS0735 jets, and those of other powerful 
radio galaxies, are created in almost 
identical oppositely directed jet and counter-jet pairs during the entire 
period of intense jet activity which may last tens of million years. 

While a pair of identical opposing jets may not always 
result in identical cavities, it is not unreasonable to expect such 
cavity pairs to be quite common, at least in their early development.  
If so, we would expect the shocks from each young cavity to collide at the 
symmetry plane perpendicular to the jet axis, reflecting as an 
additional outgoing shock similar to those shown 
in Figure (\ref{plot1}) which should be visible in deep X-ray images.
However, X-ray images of nearby clusters like Perseus 
and Virgo -- as well as more distant clusters like MS0735 -- 
often show cavities or cavity-related thermal features 
that appear to have no counterpart or misaligned counterparts 
in the opposite hemisphere. 

In particular, the image of MS0735 in \citet{mcnamara2009} shows that 
the two radio lobes and their associated X-ray cavities are not 
mirror images of each other. 
The most plausible explanation for this is that the cavities have 
been pushed aside and distorted by motions in the nearby hot cluster gas. 
We estimate from this figure that the cavities have been displaced 
transverse to their common jet axis by at least a distance of 
$d \approx 70$ kpc.
At the location of the radio cavities the sound speed in the hot 
gas is $c_s = (\gamma k_{\rm b} T / \mu m_p)^{1/2} = 513 T_{\rm keV}^{1/2}$ 
km s$^{-1} \approx 1150$ km s$^{-1}$ 
(for $T_{\rm keV} \equiv k_{\rm b} T /1\text{ keV}\approx 5$) 
and the hot gas density is $9.5 \times 10^{-27}$ gm cm$^{-3}$. 
Can the non-axisymmetric 
radio emission dislocations seen in MS0735 be generated by 
subsonic flows in the ambient gas? To create the observed spatial perturbations in MS0735, consider a ram pressure $\rho v_{\rm ram}^2$ acting on the radio lobe of area $A = \pi ab$ (where $a \sim 115$ kpc and $b \sim 70$ kpc for the northern lobe in MS0735) which accelerates the lobe by $2d/t_{\rm ram}^2 \approx \rho v_{\rm ram}^2 A/M_{\rm cav}$ in time $t_{\rm ram} \approx t_{\rm shock} \approx 10^8$ yrs. The required  gas velocity, $v_{\rm ram} \approx 9$ km s$^{-1}$ is very much less than the sound speed, suggesting that purely relativistic radio lobes are extremely vulnerable to distortion by very modest flows in the ambient cluster gas.

However, it is likely that the jet that produced 
the radio emitting cavities in MS0735 acquired 
significant masses of very hot but non-relativistic gas during 
their passage to the cavity. Suppose for simplicity that the pressure of this non-relativstic 
gas dominates in the cavities and that the cavities are in pressure equilibrium 
with the local gas, $\rho_{\rm cav}T_{\rm cav} = \rho T$. 
The ellipsoidal cavity mass is now $M_{\rm cav} = \rho_{\rm cav}(4/3) \pi a b^2$
and the ram velocity required to move the cavity a distance $d$,  
$v_{\rm ram} \approx [(8/3)(T/T_{\rm cav})bd/t_{\rm ram}^2]^{1/2}$, 
is 350 and 110 km s$^{-1}$ for $T_{\rm cav} = 50$ and 500 keV
respectively. As before, it is plausible that the cluster gas in MS0735 has 
subsonic velocities of this magnitude, 
so the lack of mirror symmetry in the radio sources tells little about 
their mass content or the almost perfect cancellation of the jet momentum that created them.
Nevertheless, a statistical study of the distortion and transverse displacement 
of radio lobes or X-ray cavities would contain information about cluster gas velocity 
fields and the baryonic component in the radio cavities.

\subsection{Energetics and Heating Effects}

Since the morphologies of the cavity and shock front in run MS-2 agree
with observations very well, we take it as our fiducial model to study
the effects of the energetic AGN outburst on the cluster MS0735 in the
rest of the paper.

\begin{figure}
\plotone{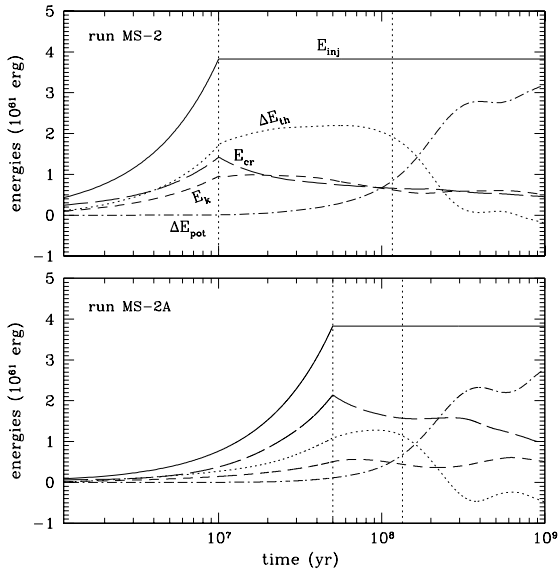}
\caption{Global energy evolution in runs MS-2 and MS-2A. The energies
  are calculated in one hemisphere and are labeled as follows: the
  injected cosmic ray energy $E_{\rm inj}$ ({\it solid}), change in
  thermal energy $\Delta E_{\rm th}$ ({\it dotted}), cosmic ray energy
  $E_{\rm cr}$ ({\it long dashed}), kinetic energy $E_{\rm k}$ ({\it
    short dashed}), and change in potential energy $\Delta E_{\rm
    pot}$ ({\it dot dashed}). The vertical lines in each panel show
  the time $t_{agn}$ when the CR injection ends (at {\it left}) and $t_{shock}$ when the shock
  propagates to $r=240$ kpc (at {\it right}), respectively.}
 \label{plot3}
 \end{figure}

The {\it top} panel of Figure \ref{plot3} shows time evolution of
global energies integrated over the whole cluster in run 
MS-2. $E_{\rm inj} (t)$ ({\it solid lines}) represents the total
injected CR energy until time $t$, which equals to the overall
increase of cluster energies, including the gas kinetic energy $E_{\rm
  k}$ ({\it short dashed lines}), the change in gas thermal energy
$\Delta E_{\rm th}$ ({\it dotted lines}), the change in gas potential
energy $\Delta E_{\rm pot}$ ({\it dot dashed}), and the total CR
energy $E_{\rm cr}$ ({\it long dashed lines}). Clearly, $E_{\rm cr}
(t)$ is much less than the total injected CR energy $E_{\rm inj} (t)$
during the CR injection phase, indicating that CRs lose a significant
fraction of their initial energy due to $pdV$ work and shock generation. At
time $t=t_{\rm agn}$ (the left dotted vertical line), more than 
$60\%$ of the injected CR energy has been converted into thermal and kinetic
energies as the thermal gas is shocked. The CR energy lost during the injection
phase is expected to be less if the same amount of energy is injected
over a longer timescale. To check this explicitly, we performed another
run MS-2A for comparison, which is the same as run MS-2, except that
the CR injection timescale is much longer ($t_{\rm agn}=50$ Myr). The
resulting energy evolution, plotted in the {\it lower} panel of
Figure \ref{plot3}, confirms that less ($\sim 50\%$) of the injected CR
energy has been converted into gas thermal and kinetic energies at
the end of the injection phase. We therefore conclude that a large
fraction ($\gtrsim 50\%$) of the AGN energy is delivered to the
ICM during the formation of X-ray cavities through shock generation
driven by $pdV$ work. This fraction increases with the CR luminosity 
$E_{agn}/t_{agn}$ during the CR injection phase.

The evolution of the gas thermal and potential energies in Figure
\ref{plot3} suggests that the thermal energy of the ICM responds to cavities in two 
competing ways. On the one hand, cavities directly heat 
the ICM through weak shocks, as clearly shown by the increase of
thermal energy with time during $t \lesssim 100$ Myr in Figure
\ref{plot3}. On the other hand, as cosmic rays displace cluster gas,
the entire cluster expands (see Fig. \ref{plot5} and
\S~\ref{section:gasmass} for more details) and thus cools as the
gravitational potential energy of the cluster increases. Local heating dominates
during the early stage of cavity evolution, while expansion cooling dominates
during the later stage when the shock becomes much weaker. 
This is indicated by the decline of $\Delta E_{\rm th}$ and the accompanying 
increase of gravitational energy during $t \gtrsim 100$ Myr in
Figure \ref{plot3}. This `global cooling' effect has been explicitly
discussed by \citet{mathews08} for a much less powerful AGN outburst
in Virgo cluster ($10^{58}$ erg); our new results for MS0735 suggest that it 
is likely to be a universal feature of AGN outbursts regardless of their energy.

\begin{figure}
\plotone{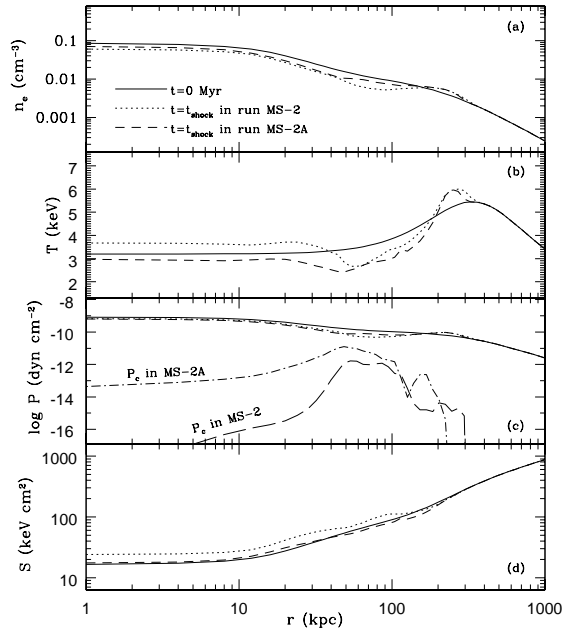}
\caption{Radial profiles of emission-weighted spherically-averaged (a) electron number density $n_{\rm{e}}$, (b) temperature $T$, (c) pressure $P$, and (d) entropy $S$ ($S\equiv k_{\rm B}T/ n_{\rm{e}}^{2/3}$) in runs MS-2 and MS-2A. The {\it long-dashed} and {\it dot-dashed} lines in panel (c) represent the cosmic ray pressure distribution in run MS-2 and MS-2A, respectively. 
}
 \label{plot4}
 \end{figure}

Figure \ref{plot4} shows radial profiles of emission-weighted
spherically-averaged gas quantities, including (a) electron number
density, (b) temperature, (c) pressure, and (d) gas entropy ($S\equiv
k_{\rm B}T/ n_{\rm{e}}^{2/3}$) in runs MS-2 and MS-2A. At $t=t_{\rm
  shock}$, the gas density in inner regions, which are already
affected by the AGN activity, drops in both models, confirming that
the cluster is undergoing global expansion. Consequently, the
expansion of these regions results in a drop in the gas temperature,
as seen in the {\it short dashed} line in Figure \ref{plot4}b for run
MS-2A. In run MS-2, the CRs are injected in a much shorter timescale,
resulting in stronger shock heating, which surpasses the effect of
global expansion at time $t_{\rm shock}$. 
Thus the central cluster gas temperature and entropy both increase in 
run MS-2 (see Fig. \ref{plot4}d). At later times,  
$t \gtrsim 2t_{\rm shock} \approx 20t_{\rm agn}$, cooling by global expansion 
begins to dominate. As explained by \citet{mathews08}, 
the expansion of cluster gas lasts much longer than the cavity since 
the volume $V$ of gas displaced by the initial cavity is approximately 
preserved after the CRs diffuse widely into the cluster gas.

\subsection{Effects on the Gas Mass Fraction}
\label{section:gasmass}

By assuming that clusters contain a proportional share of 
all cosmic baryons, cluster
gas mass fractions ($f_{\rm gas}$) have been used as a probe of the
universal ratio of baryon to total matter densities, $\Omega_{\rm
  b}/\Omega_{\rm m}$, providing one of the current best
constraints of $\Omega_{\rm m}$ when combined with accurate
determinations of $\Omega_{\rm b}$ from cosmic nucleosynthesis
calculations or observations of cosmic microwave background (CMB)
anisotropies. Furthermore, measurements of the apparent evolution of
$f_{\rm gas}$ with redshift can be used to constrain the dark
energy content of the universe (see \citealt{allen08} and references
therein). While cluster baryon mass fractions seem to provide
convincing cosmological constraints, systematic uncertainties,
especially the apparent effects of cooling and heating on the
distribution of baryons within clusters, must be well
understood and controlled (see relevant discussions in
\citealt{2009arXiv0906.4370B}). It has been noticed that cluster
baryon fractions estimated from X-ray observations are significantly
lower than the universal baryon fraction inferred from {\it WMAP} data
\citep{ettori03,vikhlinin06,mccarthy07}; in particular,
\citet{vikhlinin06} found that for a sample of $13$ Chandra clusters,
gas mass fractions within $r_{2500}$ are around $0.05$-$0.10$, a factor
of $1.7$-$3.3$ below the universal baryon fraction 
0.167 \citep{komatsu09}. Such a discrepancy
cannot be contained in the stellar/cool baryon contribution (e.g.,
\citealt{mccarthy07}). 

\begin{figure}
\plotone{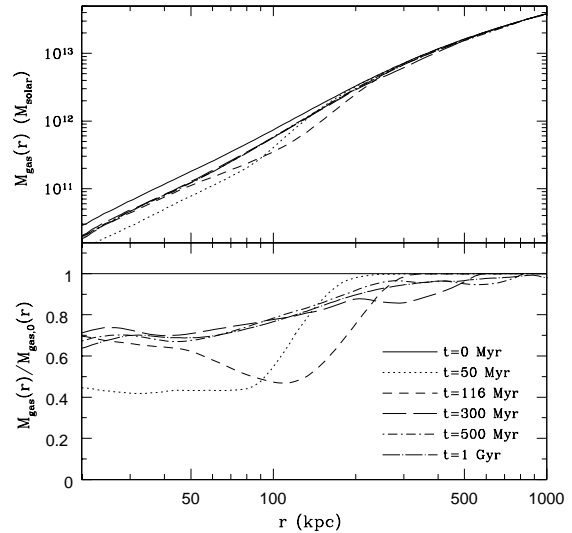}
\caption{Time evolution of the cumulative spherical gas mass within radius $r$ (top panel) and of its fractional change (bottom panel) in run MS-2. Here $M_{{\rm gas},0}(r)$ stands for $M_{\rm gas}(r)$ at $t=0$ Myr. After $t\sim 300$ Myr, the radial profile of $M_{\rm gas}(r)$ in the central regions (within $\sim 100-200$ kpc) becomes quite steady and around $20\%-30\%$ of the gas mass has been transported to large radii.
}
 \label{plot5}
 \end{figure}

However, the baryon fraction profile $f_{\rm bar}(r)$ can be 
created and shaped by AGN outbursts. 
Figure \ref{plot5} shows the time evolution of the cumulative spherical
gas mass $M_{\rm gas}(r)$ (top panel) and of its fractional change
(bottom panel) in run MS-2, where $M_{\rm gas}(r)$ is defined as the total gas
mass within a cluster-centric radius $r$. Comparing gas mass
profiles at different times, one can clearly see that, as a result of
cavity formation, the cluster undergoes a global expansion and 
gas mass is gradually transported outward. After $t\sim 300$ Myr, the
radial profile of $M_{\rm gas}(r)$ in the central regions (within
$\sim 200$ kpc) becomes quite steady and within $\sim100$ kpc
about $20\%-30\%$ of the initial gas mass has been 
transported to outer cluster regions. At $t=1$ Gyr, around
4\% of the gas mass originally within $500$ kpc,
$\sim 6 \times 10^{11}$ $M_{\odot}$, has been transported outward.
Note that, $r_{2500}\sim 465$ kpc for MS0735 \citep{gitti07}.
During the lifetime of clusters, a series of intermittent AGN outbursts are
expected to be triggered, preventing the formation of strong cooling
flows and transporting a significant fraction of the cluster gas to large
radii. This large post-cavity outflow can explain most of 
the discrepancy between the cluster baryon
mass fraction and the universal baryon fraction inferred from {\it
  WMAP} data. 
 Without correcting for this effect, systematic errors inevitably occur when using 
 $f_{\rm gas}$ at $r_{2500}$ ($r_{2500}$ is typically $\sim300$-$600$ kpc for galaxy clusters; 
 see \citealt{vikhlinin06} and \citealt{allen08}) to constrain cosmological parameters; for instance,
 $\Omega_{\rm m}$ will be systematically overestimated.  

An interesting feature of Figure \ref{plot5} is that the 
post-cavity fractional change of the gas mass decreases with radius, 
well beyond the $40 \lta r \lta 160$ kpc CR source 
region for run MS-2. 
This directly results in the trend that the gas mass fraction $f_{\rm gas}(<r)$ within a cluster-centric radius $r$
 increases with $r$, which is consistent with observations (e.g., Fig. 7 of \citealt{pratt09}).
This may also naturally explain the observational fact that the cluster gas mass
fraction within $r_{2500}$ (or $r_{500}$) increases with cluster
temperature (e.g., see Fig. 21 of \citealt{vikhlinin06}). 
For higher-temperature clusters,
$r_{2500}$ (or $r_{500}$) is usually larger, and thus a smaller fraction of the baryon gas is transported beyond
 this radius if the strength of AGN outbursts does not increase proportionally to overtake this effect. 
 On the other hand, \citet{mccarthy07} suggested that the discrepancy between the cluster baryon mass fraction 
 and the universal baryon fraction
inferred from {\it WMAP} data is mainly due to the fact that the
actual value of $\Omega_{\rm m}$ is higher than the best-fitting {\it
  WMAP} value. We argue that this explanation would result in a constant
gas mass fraction across all clusters, which is apparently
inconsistent with observations. For galaxy groups, observations show
that the gas mass fraction at $r_{2500}$ is around $0.05$, considerably 
smaller than the typical value of $\sim 0.09$ for clusters. 
The gas fraction also increases with the mean temperature in galaxy groups \citep{sun09}.

It was argued by \citet{mccarthy07} that the total energy required to produce this 
discrepancy by transporting gas mass outward may be extremely large ($\sim 10^{63}$ erg). 
However, their simple calculations assume arbitrary extrapolations of gas density profiles to 
cluster outer regions not yet observed, where the transported gas, which determines the
required increase in the gas potential energy, is located. On the other hand, our simulations
directly show that single strong AGN outbursts transport a significant amount of gas outward, 
and the increase of potential energy during the cluster expansion is accompanied by the decrease 
of thermal energy, as gas moves outward down the thermal pressure gradient (see the evolution
of thermal and potential energies in Fig. 3). Consequently, the total AGN energy needed to 
explain the gas mass fraction discrepancy, i.e., the net increase of the total cluster energy, may not
be energetically prohibitive. Further investigation of this is obviously beyond the scope of the 
current paper, and is thus left to future work.

The effect of AGN outbursts on the gas fraction has been studied with cosmological simulations of 
galaxy clusters by \citet{puchwein08}. Using a simple model for AGN mechanical heating, they find 
that AGN heating significantly reduces gas fraction within $r_{500}$ in groups and poor clusters 
while having much lesser effects on massive clusters. The analytical study by \citet{pope09} also 
suggests that AGN energy injection redistributes the ICM within the cluster central regions. These 
findings are consistent with our detailed simulations: the effect of AGN outbursts on gas fraction decreases 
with radius, and becomes quite small at $r_{500}$ for massive clusters ($r_{500}\sim 1340$ kpc 
for MS0735; \citealt{gitti07}). However, observations indicate that $f_{\rm gas}(<r)$ increases with $r$ 
at least up to $r\sim r_{500}$ \citep{pratt09}, which is roughly the largest radius for which {\it Chandra} 
and {\it XMM-Newton} data require no model extrapolation. It may suggest that the central AGN produces 
some X-ray cavities at radii much larger than what we assume here, which transport a large amount 
of gas out of $r_{500}$, and are difficult to detect due to low X-ray surface brightness in these outer regions. 
Alternatively, it is also possible that massive clusters are formed from mergers of less massive systems, 
which have already lost a significant amount of gas by AGN outbursts. Further studies are required to determine 
which scenario dominates.

\subsection{Assumptions of Hydrostatic Equilibrium and Spherical Symmetry}
\label{section:assumption}

 \begin{figure}
\plotone{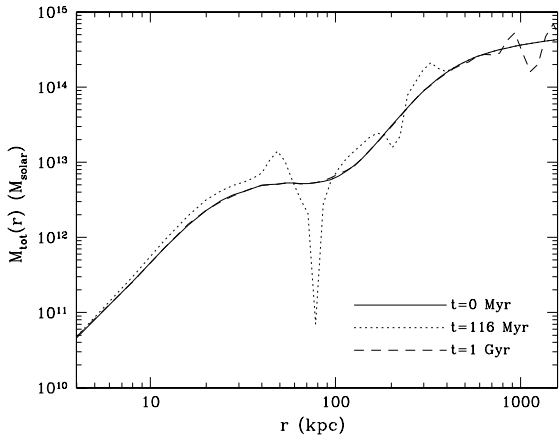}
\caption{Radial profiles of the total gravitational mass $M_{\rm tot}
  (r)$ in run MS-2 calculated at several times by assuming hydrostatic
  equilibrium and spherical symmetry. The dashed and solid lines
  largely overlap.}
 \label{plot6}
 \end{figure}

By assuming hydrostatic equilibrium and spherical symmetry, radial
profiles of cluster gas density and temperature have been widely used
to measure the total gravitational mass within radius $r$, $M_{\rm
  tot} (r)$ (e.g., \citealt{gitti07}):
\begin{eqnarray}
M_{\rm tot} (r)=-\frac{k_{\rm b}Tr}{G\mu m_{p}}\left( \frac{d\text{ ln }n_{\rm e}}{d\text{ ln }r}+\frac{d\text{ ln }T}{d\text{ ln }r}\right)\text{ ,}
\label{gmass}
\end{eqnarray}
where $G$ is the gravitational constant, $k_{\rm b}$ is the Boltzmann
constant, $m_{p}$ is the proton mass, and $\mu=0.61$ is the molecular
weight. Although this method provides one of the current best
measurements of mass profiles for relaxed clusters, the underlying
assumptions, i.e., spherical symmetry and hydrostatic equilibrium,
must be carefully tested. Systematic errors, in particular the
non-thermal pressure support from CRs or turbulent motions, have been
extensively investigated by cosmological simulations (see
\citealt{nagai07} and references therein). Here we study how X-ray
supercavities, which have yet to be included in cosmological
simulations, affect the measurements of cluster mass profiles found using 
equation (\ref{gmass}).
  
For simplicity, we ignore self-gravity of the thermal gas. Thus the
total gravitational mass, which is dominated by dark matter, does not
evolve with time in our simulations. Figure \ref{plot6} shows radial
profiles of the total gravitational mass $M_{\rm tot} (r)$ in run MS-2
calculated at various times using equation (\ref{gmass}) and the emission-weighted spherically-averaged gas density and temperature profiles (see Fig. \ref{plot4}). The
solid line in Figure \ref{plot6} represents the cluster mass profile before the AGN
outburst, while the dotted line shows the mass profile at 
$t=t_{\rm  shock} = 1.16 \times 10^8$ yrs 
when the shock propagates to $r=240$ kpc along the semiminor
axis. By comparing these two lines, one can clearly see that the total
cluster mass measured by equation (\ref{gmass}) at $t=t_{\rm shock}$
is not very accurate for $r\sim 30-300$ kpc, which is exactly
the region most strongly affected by the X-ray supercavities (see the
right-bottom panel of Fig. \ref{plot2}). As the cavity rises in the cluster atmosphere,
the ICM near the cavity is significantly disturbed, which can
be easily seen in the spherically-averaged radial temperature and
density profiles (the top two panels of Figure \ref{plot4}). Thus,
both the assumptions of hydrostatic equilibrium and spherical symmetry
are violated, and equation (\ref{gmass}) does not hold very well 
in the cavity region when $t \sim t_{\rm shock}$. 
Since $t = t_{\rm shock}$ is also the time when the deprojected gas
temperature and density profiles were observed by \citet{gitti07},
the varying slope $dM_{\rm tot}/dr$ visible in 
our initial mass distribution (solid line in Figure \ref{plot6}) 
near the cavity region also deviates somewhat from the expected Navarro-Frenk-White (NFW) form
\citep{navarro97}. This unavoidable deviation does not affect our conclusions which are all based 
on the differential changes in the cluster gas due to the cavities.
The dashed line in Figure \ref{plot6} represents the cluster mass measured by equation
(\ref{gmass}) at a much later time, $t=1$ Gyr, 
when the X-ray supercavity has already
been destroyed. At this time, the dashed line agrees very well with
the solid line in regions $r \lesssim 700$ kpc, indicating that the
cluster has returned to hydrostatic equilibrium and spherical
symmetry and equation (\ref{gmass}) provides a very accurate 
measure of the cluster mass. 
At $r \sim 1000$ kpc in Figure \ref{plot6} the dashed line oscillates
around the solid line due to outward-propagating weak
shocks produced by the outburst. 
Evidently, more accurate total mass determinations 
for MS0735 using equation (\ref{gmass}) 
could be found from observations of $n_e(r)$ and $T(r)$ in cluster quadrants 
that avoid the cavity region.

In summary, our simulations show that X-ray cavities produced by AGN
outbursts may significantly disturb the cluster gas near the cavities
and the assumptions of hydrostatic equilibrium and spherical symmetry
may thus not hold very well in these regions. This results in
systematic errors in measuring the cluster mass using
spherically-averaged profiles of gas density and temperature.  For the
powerful outburst in the cluster MS0735, the typical error in $M_{\rm tot}(r)$ at 
radii near the cavities is around $10-30\%$, but may be much
larger in some local regions. We have also done similar calculations with
much less energetic AGN outbursts, and found that the cavities created 
are smaller and thus affect smaller cluster regions. X-ray
cavities do not affect the measurements of cluster mass in the regions
far away from them, in particular, in the cluster outer regions (e.g.,
$r_{500}$).

\section{Summary and Discussion}
\label{section:conclusion}

By conducting a suite of two-dimensional axisymmetric hydrodynamical
simulations, we investigated the formation and evolution of the X-ray
supercavities recently observed in the cluster MS0735, which is the
most energetic AGN outburst known. We assume that the X-ray
supercavities are inflated by cosmic rays injected from AGN jets, and
follow the co-evolution of the cluster gas and the CRs. We then study
the thermal and hydrodynamic effects of this energetic AGN outburst on
the cluster MS0735. Here we briefly summarize our main findings:

1. X-ray deficient cavities and weak shocks are successfully created
by AGN-generated cosmic rays. In particular, run MS-2 reproduces very
well many characteristic observational features of the cluster MS0735,
such as the size and location of the cavities, the location and
strength of the shock.  In this run, the age and total energy of this AGN
outburst are $116$ Myr and $7.6\times10^{61}$ erg, consistent with
previous estimates \citep{mcnamara05, mcnamara2009}.

2. Assuming that X-ray cavities are generated by 
relativistic AGN jets, the jet that produced the north cavity in run MS-2 had a 
momentum $\sim 2 \times 10^7 M_{\odot}c$ which must be balanced
by an almost identical southerly jet ejection. Otherwise, 
a one-sided jet in MS0735 would cause its AGN black hole 
to attain a recoil velocity of $\sim 6000(M_{\rm bh}/10^{9}M_{\sun})^{-1}$ km s$^{-1}$, 
which is large enough to completely 
eject a massive black hole from the center of even 
the most massive elliptical galaxies and their surrounding group or cluster dark matter halos 
\citep{merritt04}. 
The radio emission from the cavities in MS0735, while clearly bipolar, 
is not perfectly symmetric. 
This could be due to the production of somewhat different cavities, 
perhaps at different cluster radii,  by two jets that were identical 
when they left the AGN. 
In addition, we show that small subsonic motions in the hot gas of MS0735 are 
sufficient to explain the axisymmetric cavity radio and X-ray emission observed.
However, if the cavities formed by the two opposing jets were initially 
approximately equal in power and formed at similar distances from the cluster 
center, the shocks expanding away from each cavity would collide at the 
symmetry plane, producing a second, somewhat weaker shock in each 
hemisphere as seen in our Figure \ref{plot1}. 
Efforts should be made to detect these symmetry shocks in deep X-ray images 
since they contain information about the identical nature of the two jets and 
their cavities.

3. Run MS-1, where CRs are injected at a fixed location, produces an
oblate, radially-flattened X-ray cavity unlike the observed cavity which is 
radially elongated. In contrast, in run MS-2 which fits observations very well,
the CRs are injected continuously along the jet direction as it moves
away from the cluster center. The radial elongation of the cavity may
also be produced if the CRs are injected at a series of locations
along the jet direction.

4. During the CR injection phase, more than half of the injected CR
energy is converted into thermal and kinetic energy of the
ICM. The CRs lose energy thorough $pdV$ work and shock generation as
they displace the cluster gas. For instance, $\sim 63\%$ of the injected CR energy has been lost at $t=t_{\rm agn} = 10^7$ yrs in run MS-2. The fraction of initial CR energy lost
at time $t_{\rm agn}$ increases with the CR luminosity during the injection phase.

5. The thermal energy of the ICM responds to CR-generated cavities in several complex ways. As the cavities are inflated in MS0735, shocks are driven into the surrounding gas, heating the local ICM. Mixing by CR buoyancy may  also increase the gas entropy near the post-cavity site as relatively low entropy, CR-enriched gas near the cluster core flows outward with the CRs and is replaced by inflowing gas of higher-entropy. On the other hand, the production of new cavities causes the cluster gas to readjust outwards in the cluster, increasing its global potential energy and decreasing its thermal energy (cooling) as its density decreases. Local heating by cavity shocks dominates during the early stage of cavity evolution, whereas cooling by global expansion dominates at later times, when the total cluster thermal energy decreases below the original cluster thermal energy before the outburst. Thus, from a global perspective, the cluster gas is eventually cooled by the cavity, as previously shown by \citet{mathews08} for a low-energy outburst in the Virgo cluster. We have now confirmed this for the cluster MS0735 which is undergoing the most powerful AGN outburst known. This ultimate allocation of cavity energy is likely to be universal for all AGN outbursts.

6. The creation of X-ray supercavities in MS0735 
with CRs produces a huge, long-lasting outward migration of cluster
gas to large radii. This mass outflow arises because of the global cluster expansion due to 
the gas displaced as the cavity forms and 
heating by outward-propagating shock waves driven by the expanding cavity.
The mass outflow following the single energetic outburst in MS073 
is large. Repeated AGN outbursts may naturally explain why the cluster baryon mass fraction 
in MS0735 and other similar clusters is much less than the
universal baryon fraction inferred from {\it WMAP} data, and why it increases
with the cluster temperature.

7. X-ray supercavities significantly disturb the cluster gas near
them, locally invalidating the assumptions of hydrostatic equilibrium
and spherical symmetry. This results in systematic errors (typically around $10-30\%$ for MS0735) in measuring the cluster mass in these regions using spherically-averaged gas density and temperature profiles.

8. Assuming that X-ray cavities are filled with relativistic CRs, $4PV$ is often used
to estimate the energy released by each AGN jet. 
When combined with a typical (buoyant) cavity lifetime $t_{\rm life}$, 
the jet power $4PV/t_{\rm life}$ can be estimated and compared with the 
bolometric X-ray luminosity of the cluster gas \citep{birzan04}. 
Since the total CR energy contained inside the cavity is $E_{\rm cr}V=3PV$, where $E_{\rm cr}$ is the average CR energy density in the cavity, about $1/4$ of the jet
energy ($4PV-E_{\rm cr}V= PV$) is delivered to the ICM during the cavity formation in these
approximations. In our simulations, we directly studied the energetics of
the cavity, and find that more than half of the injected CR energy 
goes into the ICM, i.e., the CR energy lost during the cavity formation 
is actually more than $3PV$. This fraction increases with the CR luminosity.
Although $4PV$ provides an approximate estimation for AGN jet energy, the
actual AGN energy may be a few times larger (typically $\sim 6PV-10PV$). 
This is consistent with previous simulations of cavities produced by (non-relativistic) thermal jets \citep{binney07}, which, due to the additional inclusion of a large amount of kinetic energy in thermal jets, 
find that AGN energy is around $15 PV$, even larger than our estimation. 
Our calculations further show that $4PV$ begins to decrease after the cavity formation as CRs diffuse through the cavity walls \citep{mathews08}. More generally, if $4PV/t_{\rm life}$ underestimates the cluster jet power, 
it becomes easier to match this heating with the bolometric radiative losses 
$L_X$ as proposed by \citet{birzan04} and \citet{rafferty06}.

We ignore direct interactions of CRs with the ambient cluster gas,
e.g., Coulomb interactions, hadronic collisions and interactions
through the generation of hydromagnetic waves, which may become
important especially when the cavities are disrupted and when the CRs
are mixed with the ICM \citep{2008MNRAS.384..251G}. These interactions
may provide significant heating effects for the thermal gas. One
byproduct of these interactions is the generation of $\gamma
\text{-rays}$ through the decay of neutral pions, which may be studied
by the recently-launched {\it Fermi} telescope \citep{ando08, mathews09}.
Previous cluster observations with the High Energy Stereoscopic System (HESS) 
have not detected these $\gamma \text{-ray}$ signals in the VHE ($> 100$ GeV) band,
providing upper limits on $\gamma \text{-ray}$ fluxes and ratios of cosmic ray pressure to 
thermal pressure in galaxy clusters (e.g., \citealt{domainko09}).

The center of each cavity in the cluster MS0735 is located at a projected 
cluster-centric radius of around $170$ kpc, which is much larger than
the average projected cluster radius of $20$ kpc for X-ray cavities typically observed in
clusters \citep{birzan04}. If strong AGN outbursts 
having energies similar to that of MS0735 create X-ray
cavities much closer to the cluster center than those observed in
MS0735, they produce much stronger heating in the cluster cool core
and may even fully destroy it, transforming a cool core cluster to a
non-cool core cluster \citep{guo09}. It is of great interest to investigate if metallicity peaks at cluster
centers are also destroyed in this process, producing much flatter
metallicity profiles as observed in non-cool core systems (e.g.,
\citealt{de-grandi01}; \citealt{baldi07}). We will thoroughly
investigate these models in a follow-up paper.

\acknowledgements

We thank Fabrizio Brighenti for useful discussions, 
and the anonymous referee for a positive and helpful report. 
Studies of the evolution of hot cluster gas at UC Santa Cruz are supported by NSF and NASA grants for which we are very grateful.

\bibliography{ms}

\begin{thebibliography}{48}
\expandafter\ifx\csname natexlab\endcsname\relax\def\natexlab#1{#1}\fi

\bibitem[{{Allen} {et~al.}(2008){Allen}, {Rapetti}, {Schmidt}, {Ebeling},
  {Morris}, \& {Fabian}}]{allen08}
{Allen}, S.~W., {Rapetti}, D.~A., {Schmidt}, R.~W., {Ebeling}, H., {Morris},
  R.~G., \& {Fabian}, A.~C. 2008, \mnras, 383, 879

\bibitem[{{Ando} \& {Nagai}(2008)}]{ando08}
{Ando}, S., \& {Nagai}, D. 2008, \mnras, 385, 2243

\bibitem[{{Baldi} {et~al.}(2007){Baldi}, {Ettori}, {Mazzotta}, {Tozzi}, \&
  {Borgani}}]{baldi07}
{Baldi}, A., {Ettori}, S., {Mazzotta}, P., {Tozzi}, P., \& {Borgani}, S. 2007,
  \apj, 666, 835

\bibitem[{{Binney} {et~al.}(2007){Binney}, {Alouani Bibi}, \&
  {Omma}}]{binney07}
{Binney}, J., {Alouani Bibi}, F., \& {Omma}, H. 2007, \mnras, 377, 142

\bibitem[{{B{\^i}rzan} {et~al.}(2004){B{\^i}rzan}, {Rafferty}, {McNamara},
  {Wise}, \& {Nulsen}}]{birzan04}
{B{\^i}rzan}, L., {Rafferty}, D.~A., {McNamara}, B.~R., {Wise}, M.~W., \&
  {Nulsen}, P.~E.~J. 2004, \apj, 607, 800

\bibitem[{{Borgani} \& {Kravtsov}(2009)}]{2009arXiv0906.4370B}
{Borgani}, S., \& {Kravtsov}, A. 2009, preprint (arXiv:0906.4370)

\bibitem[{{Br{\"u}ggen} {et~al.}(2007){Br{\"u}ggen}, {Heinz}, {Roediger},
  {Ruszkowski}, \& {Simionescu}}]{brueggen07}
{Br{\"u}ggen}, M., {Heinz}, S., {Roediger}, E., {Ruszkowski}, M., \&
  {Simionescu}, A. 2007, \mnras, 380, L67

\bibitem[{{Br{\"u}ggen} {et~al.}(2009){Br{\"u}ggen}, {Scannapieco}, \&
  {Heinz}}]{brueggen09a}
{Br{\"u}ggen}, M., {Scannapieco}, E., \& {Heinz}, S. 2009, \mnras, 395, 2210

\bibitem[{{Campanelli} {et~al.}(2007){Campanelli}, {Lousto}, {Zlochower}, \&
  {Merritt}}]{campanelli07}
{Campanelli}, M., {Lousto}, C.~O., {Zlochower}, Y., \& {Merritt}, D. 2007,
  Physical Review Letters, 98, 231102

\bibitem[{{De Grandi} \& {Molendi}(2001)}]{de-grandi01}
{De Grandi}, S., \& {Molendi}, S. 2001, \apj, 551, 153

\bibitem[{{Dolag} {et~al.}(2001){Dolag}, {Schindler}, {Govoni}, \&
  {Feretti}}]{dolag01}
{Dolag}, K., {Schindler}, S., {Govoni}, F., \& {Feretti}, L. 2001, \aap, 378,
  777

\bibitem[{{Domainko} {et~al.}(2009){Domainko}, {Nedbal}, {Hinton}, \&
  {Martineau-Huynh}}]{domainko09}
{Domainko}, W., {Nedbal}, D., {Hinton}, J.~A., \& {Martineau-Huynh}, O. 2009,
  International Journal of Modern Physics D, 18, 1627

\bibitem[{{Ettori}(2003)}]{ettori03}
{Ettori}, S. 2003, \mnras, 344, L13

\bibitem[{{Fabian} {et~al.}(2006){Fabian}, {Sanders}, {Taylor}, {Allen},
  {Crawford}, {Johnstone}, \& {Iwasawa}}]{fabian06}
{Fabian}, A.~C., {Sanders}, J.~S., {Taylor}, G.~B., {Allen}, S.~W., {Crawford},
  C.~S., {Johnstone}, R.~M., \& {Iwasawa}, K. 2006, \mnras, 366, 417

\bibitem[{{Ferrarese} \& {Merritt}(2000)}]{ferrarese00}
{Ferrarese}, L., \& {Merritt}, D. 2000, \apjl, 539, L9

\bibitem[{{Gebhardt} {et~al.}(2000){Gebhardt}, {Bender}, {Bower}, {Dressler},
  {Faber}, {Filippenko}, {Green}, {Grillmair}, {Ho}, {Kormendy}, {Lauer},
  {Magorrian}, {Pinkney}, {Richstone}, \& {Tremaine}}]{gebhardt00}
{Gebhardt}, K., {et~al.} 2000, \apjl, 539, L13

\bibitem[{{Gitti} {et~al.}(2007){Gitti}, {McNamara}, {Nulsen}, \&
  {Wise}}]{gitti07}
{Gitti}, M., {McNamara}, B.~R., {Nulsen}, P.~E.~J., \& {Wise}, M.~W. 2007,
  \apj, 660, 1118

\bibitem[{{Guo} \& {Oh}(2008)}]{2008MNRAS.384..251G}
{Guo}, F., \& {Oh}, S.~P. 2008, \mnras, 384, 251

\bibitem[{{Guo} \& {Oh}(2009)}]{guo09}
---. 2009, \mnras, 400, 1992

\bibitem[{{Guo} {et~al.}(2008){Guo}, {Oh}, \& {Ruszkowski}}]{guo08b}
{Guo}, F., {Oh}, S.~P., \& {Ruszkowski}, M. 2008, \apj, 688, 859

\bibitem[{{Komatsu} {et~al.}(2009){Komatsu}, {Dunkley}, {Nolta}, {Bennett},
  {Gold}, {Hinshaw}, {Jarosik}, {Larson}, {Limon}, {Page}, {Spergel},
  {Halpern}, {Hill}, {Kogut}, {Meyer}, {Tucker}, {Weiland}, {Wollack}, \&
  {Wright}}]{komatsu09}
{Komatsu}, E., {et~al.} 2009, \apjs, 180, 330

\bibitem[{{Madau} \& {Quataert}(2004)}]{madau04}
{Madau}, P., \& {Quataert}, E. 2004, \apjl, 606, L17

\bibitem[{{Mathews}(2009)}]{mathews09}
{Mathews}, W.~G. 2009, \apjl, 695, L49

\bibitem[{{Mathews} \& {Brighenti}(2008{\natexlab{a}})}]{mathews08a}
{Mathews}, W.~G., \& {Brighenti}, F. 2008{\natexlab{a}}, \apj, 676, 880

\bibitem[{{Mathews} \& {Brighenti}(2008{\natexlab{b}})}]{mathews08}
---. 2008{\natexlab{b}}, \apj, 685, 128

\bibitem[{{Mathews} {et~al.}(2006){Mathews}, {Faltenbacher}, \&
  {Brighenti}}]{mathews06}
{Mathews}, W.~G., {Faltenbacher}, A., \& {Brighenti}, F. 2006, \apj, 638, 659

\bibitem[{{McCarthy} {et~al.}(2007){McCarthy}, {Bower}, \&
  {Balogh}}]{mccarthy07}
{McCarthy}, I.~G., {Bower}, R.~G., \& {Balogh}, M.~L. 2007, \mnras, 377, 1457

\bibitem[{{McNamara} {et~al.}(2009){McNamara}, {Kazemzadeh}, {Rafferty},
  {B{\^i}rzan}, {Nulsen}, {Kirkpatrick}, \& {Wise}}]{mcnamara2009}
{McNamara}, B.~R., {Kazemzadeh}, F., {Rafferty}, D.~A., {B{\^i}rzan}, L.,
  {Nulsen}, P.~E.~J., {Kirkpatrick}, C.~C., \& {Wise}, M.~W. 2009, \apj, 698,
  594

\bibitem[{{McNamara} \& {Nulsen}(2007)}]{2007ARA&A..45..117M}
{McNamara}, B.~R., \& {Nulsen}, P.~E.~J. 2007, \araa, 45, 117

\bibitem[{{McNamara} {et~al.}(2005){McNamara}, {Nulsen}, {Wise}, {Rafferty},
  {Carilli}, {Sarazin}, \& {Blanton}}]{mcnamara05}
{McNamara}, B.~R., {Nulsen}, P.~E.~J., {Wise}, M.~W., {Rafferty}, D.~A.,
  {Carilli}, C., {Sarazin}, C.~L., \& {Blanton}, E.~L. 2005, \nat, 433, 45

\bibitem[{{Merritt} {et~al.}(2004){Merritt}, {Milosavljevi{\'c}}, {Favata},
  {Hughes}, \& {Holz}}]{merritt04}
{Merritt}, D., {Milosavljevi{\'c}}, M., {Favata}, M., {Hughes}, S.~A., \&
  {Holz}, D.~E. 2004, \apjl, 607, L9

\bibitem[{{Nagai} {et~al.}(2007){Nagai}, {Vikhlinin}, \& {Kravtsov}}]{nagai07}
{Nagai}, D., {Vikhlinin}, A., \& {Kravtsov}, A.~V. 2007, \apj, 655, 98

\bibitem[{{Navarro} {et~al.}(1997){Navarro}, {Frenk}, \& {White}}]{navarro97}
{Navarro}, J.~F., {Frenk}, C.~S., \& {White}, S.~D.~M. 1997, \apj, 490, 493

\bibitem[{{Peterson} \& {Fabian}(2006)}]{2006PhR...427....1P}
{Peterson}, J.~R., \& {Fabian}, A.~C. 2006, \physrep, 427, 1

\bibitem[{{Peterson} {et~al.}(2003){Peterson}, {Kahn}, {Paerels}, {Kaastra},
  {Tamura}, {Bleeker}, {Ferrigno}, \& {Jernigan}}]{2003ApJ...590..207P}
{Peterson}, J.~R., {Kahn}, S.~M., {Paerels}, F.~B.~S., {Kaastra}, J.~S.,
  {Tamura}, T., {Bleeker}, J.~A.~M., {Ferrigno}, C., \& {Jernigan}, J.~G. 2003,
  \apj, 590, 207

\bibitem[{{Peterson} {et~al.}(2001){Peterson}, {Paerels}, {Kaastra}, {Arnaud},
  {Reiprich}, {Fabian}, {Mushotzky}, {Jernigan}, \&
  {Sakelliou}}]{2001A&A...365L.104P}
{Peterson}, J.~R., {et~al.} 2001, \aap, 365, L104

\bibitem[{{Pope}(2009)}]{pope09}
{Pope}, E.~C.~D. 2009, \mnras, 395, 2317

\bibitem[{{Pratt} {et~al.}(2009){Pratt}, {Arnaud}, {Piffaretti}, {Boehringer},
  {Ponman}, {Croston}, {Voit}, {Borgani}, \& {Bower}}]{pratt09}
{Pratt}, G.~W., {et~al.} 2009, preprint (arXiv:0909.3776)

\bibitem[{{Puchwein} {et~al.}(2008){Puchwein}, {Sijacki}, \&
  {Springel}}]{puchwein08}
{Puchwein}, E., {Sijacki}, D., \& {Springel}, V. 2008, \apjl, 687, L53

\bibitem[{{Rafferty} {et~al.}(2006){Rafferty}, {McNamara}, {Nulsen}, \&
  {Wise}}]{rafferty06}
{Rafferty}, D.~A., {McNamara}, B.~R., {Nulsen}, P.~E.~J., \& {Wise}, M.~W.
  2006, \apj, 652, 216

\bibitem[{{Ruszkowski} {et~al.}(2004){Ruszkowski}, {Br{\"u}ggen}, \&
  {Begelman}}]{ruszkowski04}
{Ruszkowski}, M., {Br{\"u}ggen}, M., \& {Begelman}, M.~C. 2004, \apj, 611, 158

\bibitem[{{Ruszkowski} {et~al.}(2007){Ruszkowski}, {En{\ss}lin}, {Br{\"u}ggen},
  {Heinz}, \& {Pfrommer}}]{ruszkowski07}
{Ruszkowski}, M., {En{\ss}lin}, T.~A., {Br{\"u}ggen}, M., {Heinz}, S., \&
  {Pfrommer}, C. 2007, \mnras, 378, 662

\bibitem[{{Sternberg} \& {Soker}(2009)}]{soker09}
{Sternberg}, A., \& {Soker}, N. 2009, \mnras, 398, 422

\bibitem[{{Stone} \& {Norman}(1992)}]{stone92}
{Stone}, J.~M., \& {Norman}, M.~L. 1992, \apjs, 80, 753

\bibitem[{{Sun} {et~al.}(2009){Sun}, {Voit}, {Donahue}, {Jones}, {Forman}, \&
  {Vikhlinin}}]{sun09}
{Sun}, M., {Voit}, G.~M., {Donahue}, M., {Jones}, C., {Forman}, W., \&
  {Vikhlinin}, A. 2009, \apj, 693, 1142

\bibitem[{{Sutherland} \& {Dopita}(1993)}]{1993ApJS...88..253S}
{Sutherland}, R.~S., \& {Dopita}, M.~A. 1993, \apjs, 88, 253

\bibitem[{{Tamura} {et~al.}(2001){Tamura}, {Kaastra}, {Peterson}, {Paerels},
  {Mittaz}, {Trudolyubov}, {Stewart}, {Fabian}, {Mushotzky}, {Lumb}, \&
  {Ikebe}}]{2001A&A...365L..87T}
{Tamura}, T., {et~al.} 2001, \aap, 365, L87

\bibitem[{{Vikhlinin} {et~al.}(2006){Vikhlinin}, {Kravtsov}, {Forman}, {Jones},
  {Markevitch}, {Murray}, \& {Van Speybroeck}}]{vikhlinin06}
{Vikhlinin}, A., {Kravtsov}, A., {Forman}, W., {Jones}, C., {Markevitch}, M.,
  {Murray}, S.~S., \& {Van Speybroeck}, L. 2006, \apj, 640, 691

\end{thebibliography}

\label{lastpage}

\end{document}